\documentclass[conference]{IEEEtran}

\usepackage{filecontents}

\makeatletter
\def\endthebibliography{%
  \def\@noitemerr{\@latex@warning{Empty `thebibliography' environment}}%
  \endlist
}
\makeatother
\IEEEoverridecommandlockouts
\usepackage{cite}
\usepackage{amsmath,amssymb,amsfonts}
\usepackage{algorithmic}
\usepackage{graphicx}
\usepackage{textcomp}
\usepackage{xcolor}
\usepackage{flushend}
\usepackage{fourier} 
\usepackage{array}
\usepackage{makecell}

\def\BibTeX{{\rm B\kern-.05em{\sc i\kern-.025em b}\kern-.08em
    T\kern-.1667em\lower.7ex\hbox{E}\kern-.125emX}}
\begin{document}

\title{DeepAcid: Classification of macromolecule type based on sequences of amino acids
\thanks{}
}

\author{
    \IEEEauthorblockN{Sarwar Khan$^{1,2,3}$ \\ $^1 $National Chengchi University\\ $^2 $Taiwan international graduate program\\$^3$Institute of information science academia Sinica.}
   
}

\maketitle

\begin{abstract}
The study of the amino acid sequence is vital in life sciences. In this paper, we are using deep learning to solve macromolecule classification problem using amino acids. Deep learning has emerged as a strong and efficient framework that can be applied to a broad spectrum of complex learning problems which were difficult to solve using traditional machine learning techniques in the past. We are using word embedding from NLP to represent the amino acid sequence as vectors. We are using different deep learning model for classification of macromolecules like CNN, LSTM, and GRU. Convolution neural network can extract features from amino acid sequences which are represented by vectors. The extracted features will be feed to a different type of model to train a robust classifier. our results show that Word2vec as embedding combine with VGG-16 has better performance than LSTM and GRU. our approach gets an error rate of 1.5\%. Code is available at $https://github.com/say2sarwar/DeepAcid$ \\
\end{abstract}


\section{Introduction}
The last decade has witnessed the great success of deep learning as it has brought revolutionary advances in many application domains, including computer vision, natural language processing, and signal processing. The key idea behind deep learning is to consider feature learning and classification in the same network architecture, and use back-propagation to update model parameters to learn discriminative feature representations. More importantly, many novel deep learning methods have been devised and improved classification performance significantly~\cite{he2016deep, liu2019time, szegedy2016rethinking}.

Lee et al.~\cite{lee2016protein} targeted on learning an informative feature representation of protein sequence as the input of neural network models to obtain the final predicting output of belonging protein family. Hou et al.~\cite{hou2017deepsf} proposed a framework with deep 1D CNN (DeepSF) which is robust on both fold recognition and the study of sequence-structure relationship to classify protein sequence. Nguyen et al.~\cite{nguyen2016dna} developed a framework with convolution neural network which used the idea of translation to convert DNA sequence to word sequence as for final classification. 

The revolution in machine learning particularly deep learning ~\cite{lecun2015deep,kriz2012,lenet1998} made it possible to study and extract a complex pattern from data in order to make the machine model more robust. Study of DNA in life sciences in an important factor to understand organisms. Current sequencing technologies made it possible to read DNA sequences with lower cost. DNA databases are increasing day by day and we need to use the power of modern computing to help understand the DNA. one of the most important and basic tasks is to classify DNA sequences. 

This work focus on the classification of macromolecule based on amino acids sequences. Within all lifeforms on Earth, from the tiniest bacterium to the giant sperm whale, there are four major classes of organic macromolecules that are always found and are essential to life.  These are the carbohydrates, lipids (or fats), proteins, and nucleic acids.  All of the major macromolecule classes are similar, in that, they are large polymers that are assembled from small repeating monomer subunits. Proteins are large, complex molecules that play many critical roles in the body. They are made up of hundreds or thousands of smaller units called amino acids, which are attached to one another in long chains. There are 20 different types of amino acids that can be combined to make a protein. The name of these 20 common amino acids are as follows: alanine, arginine, asparagine, aspartic acid, cysteine, glutamic acid, glutamine, glycine, histidine, isoleucine, leucine, lysine, methionine, phenylalaine, proline, serine, threonine, tryptophan, tyrosine, and valine. The sequence of amino acids determines each protein's unique 3-dimensional structure and its specific function. 

Carbohydrates are polymers that include both sugars and polymers of sugars, and they serve as fuel and building materials both within and outside of the cells. For instance, fructose and glucose are examples of carbohydrates which are essential to life. Nucleic acids are polymeric macromolecules that are essential for all known forms of life. The two types of nucleic acids are DNA and RNA, which are both found in nuclei of cells. They allow organisms to reproduce their complex components.

\section{Problem statement}
The interaction of Protein with protein and protein with DNA/RNA play a pivotal role in protein function. Experimental detection of residues in protein-protein interaction surfaces must come from the determination of the structure of protein-protein, protein-DNA, and protein-RNA complexes. However, experimental determination of such complexes lags far behind the number of known protein sequences. 
Hence, there is a need for the development of reliable computational methods for identifying protein-protein, protein-RNA, and protein-DNA interface residues. Identification of macromolecules and detection of specific amino acid residues that contribute to the strength of interactions is an important problem with broad applications ranging from rational drug design to the analysis of metabolic and signal transduction networks. Against this background, this project is aimed at developing a machine learning algorithm that identifies the macros molecule types given the sequence of amino acid, and residue count.

\section{MATERIAL AND METHOD}
\subsection{Dataset}
\begin{figure}[t]
\includegraphics[width=\linewidth]{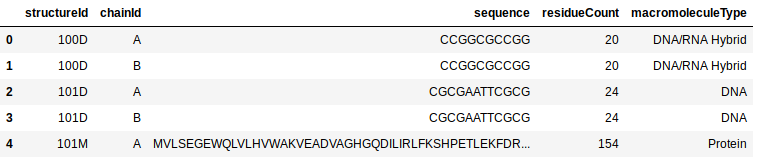}
\caption{PDB data sequence}
\label{fig:data1}
\end{figure}

The dataset contains two files with a different number of entries.  Figure \ref{fig:data1} shows the first five rows of the file. The dataset has 467304 entries with five columns. Table \ref{table:data1} shows the database description. \\
As we can see from the Table \ref{table:data1} that we have 4 types of macromolecule Protein, DNA, RNA and Protein/DNA/RNA Hybrid. We dropped the other types during the pre-processing step. The second file is also arrange based on structureId. This file contains protein meta-data i.e. resolution, extraction method, experimental technique, etc. this file has 141401 entries with 14 columns. We can merge both files, based structureId . the very step of pre-processing is to drop all the entries with NaN value or if label or sequence is missing. After removing the missing values, the sequence is checked for tags or numbers and remove it. once the data is cleaned. we divide the sequence into tri-gram, each sequence in now a combination of three characters strings. e.g. (CGC GAA TTC GCG). the final block may not have all three and we added 0 in order to make it equal slice (padding). \\
The final output contains two columns, one for sequence and the other for the label. there is 432474 rows in the processed data with four label as discussed earlier. This is unbalanced data and we will discuss the augmentation incoming section.

We also create a special case of dataset to balance and normalize all the classes. we take 424 sequences of each class to create mini dataset and test the performance of all the models. This mini dataset have almost the same results and the whole set with some down/up sampling.

\begin{table}[htbp]
\caption{Dataset description.}
\label{table:data1}
\begin{tabular}{ |c|c|c|c| } 

 \hline
 \thead{Label} & \thead{TYPE} & \thead{Data structure} & \thead{Unique \\ Entries} \\ 
 
 \hline
\thead{structureId}  & \thead{structure ID} & \thead{object} & \thead{140250} \\ 
 
 \hline
 \thead{chainId} & \thead{Chain ID} & \thead{object} & \thead{2837} \\ 
 
 \hline
 \thead{sequence} & \thead{Protein sequence} & \thead{Object} & \thead{104813} \\
 
 \hline
 \thead{residue Count} & \makecell{No of residues \\ (ATCG's)} & \thead{Integer} & \thead{4737} \\
 
 \hline
\makecell{macromolecule \\ Type} & \makecell{Type of \\ Macro-molecule} & \thead{Object} & \thead{14} \\

 \hline
\end{tabular}

\end{table}

\subsection{Biological Structures in the Dataset }
DNA makes RNA, RNA makes amino-acids, amino-acid makes protein. This is known as the central dogma of life. A DNA or RNA is made of Nucleotides, which are of four types (A, T/U, C, G). The nucleotide sequence is the combination of these nucleotides in a row. 
Three nucleotides combine to form codon which is building a block of Amino Acids. The amino-acid then combines to form proteins. To make a protein at least 20 amino acids are necessary. \newline
Let's explain it with a real example. 
ATT is a codon, which is basically three nucleotides. This codon represents amino-acid (isoleucine) represented by the letter "I". TTT is another codon which represents another amino-acid ( phenylalanine) and is represented by letter "F". These "IF" combines along with others to make protein. The letters in codon represent nucleotides while the letters in protein sequence represent amino acids.\newline At least 20 amino-acid must combine to make one functional protein. The maximum number depends on when machinery overcome a stop codon to stop making one protein. The machinery may overcome a stop codon after 20 or may overcome after 500. the amino acid sequence determines the type of proteins.

\begin{figure*}[t]
\includegraphics[width=\linewidth]{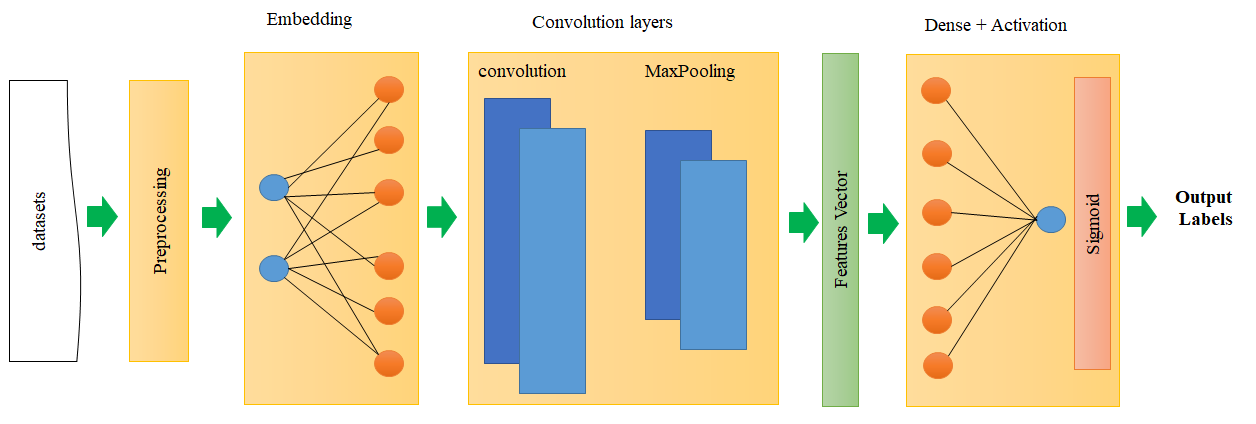}
\caption{Proposed method using embedding and CNN.}
\label{fig:model1}
\end{figure*}
\begin{figure}[ht]
\includegraphics[width=\linewidth]{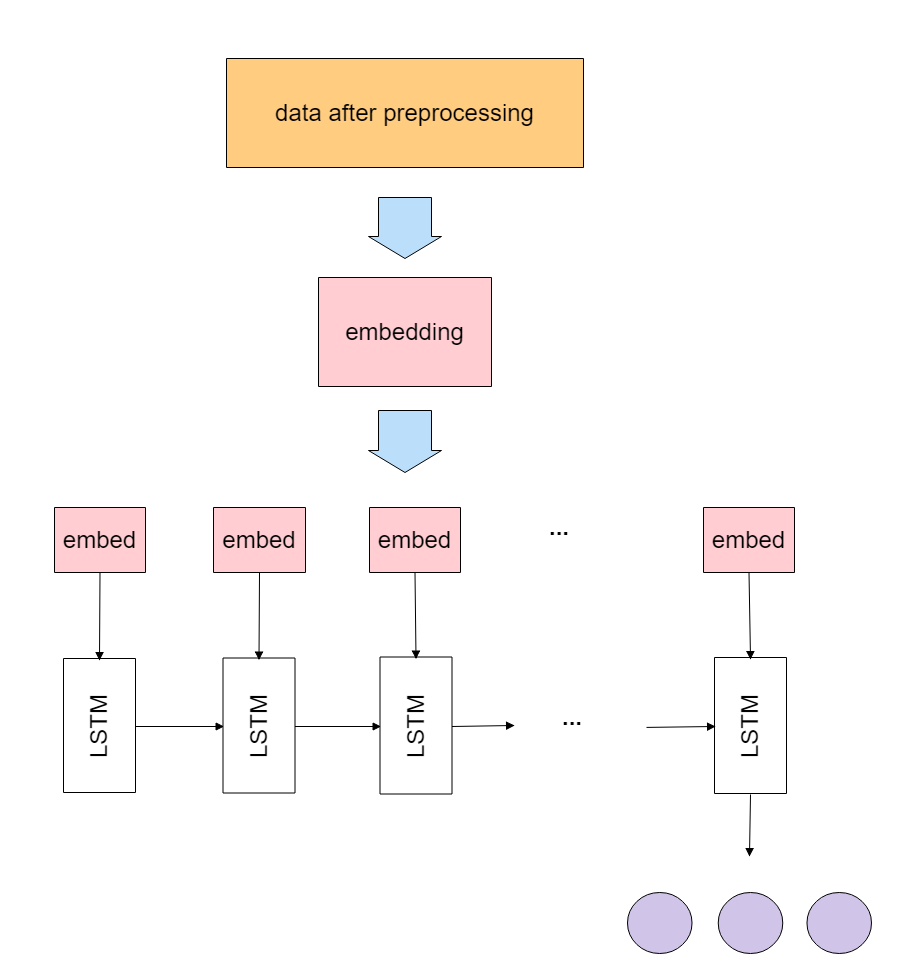}
\caption{LSTM model using embedding at input}
\label{fig:LSTM}
\end{figure}
\begin{figure}[ht]
\includegraphics[width=\linewidth]{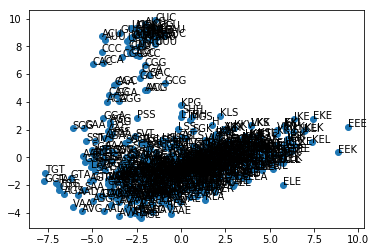}
\caption{word2vec model relation between sequences}
\label{fig:word2vec}
\end{figure}

\section{Proposed Method}
The task of classifying macromolecule type using sequence can be seen as a sequence classification problem. This is analogous to the sentence classification task in NLP. Thus, we apply a skip-gram analysis from NLP research to model our problem. There are various sequence models in the deep learning domain. Among them, we used LSTM, GRU and 1D Convolution for our problem. Recurrent neural networks, such as the Long Short-Term Memory, or LSTM, network are specifically designed to support sequences of input data. \ref{fig:LSTM} shows the layout of the model. 

They are capable of learning the complex dynamics within the temporal ordering of input sequences as well as use internal memory to remember or use information across long input sequences.

As it is crucial to any deep learning task, we applied a data prepossessing task before feeding to our model. This task includes handling missing values, downsampling dominant class to balance the distribution of data.

Figure \ref{fig:model1} shows the block diagram of the proposed system. This system can be broadly divided into three categories. first is dataset processing, the second part of this model is embedding. we have different choice in embedding but word2vec ~\cite{word2vec}, Fast-Text ~\cite{fasttext} and GloVe ~\cite{Glove} are famous embedding technique in NLP. these embedding has been tested in different bioinformatics task and the results are promising. we used  word2vec in this task. one-hot vector is another famous embedding method for amino acid representation and we used it for comparison to other embedding technique.

The final category is CNN, we will have to determine the number of layers this network will need. we also need to find hyper-parameters along with size and height and width of the model. the output layer is a softmax layer is used to classify the sequence. the coming section explains these parts in details.

\begin{figure}[t]
\includegraphics[width=\linewidth]{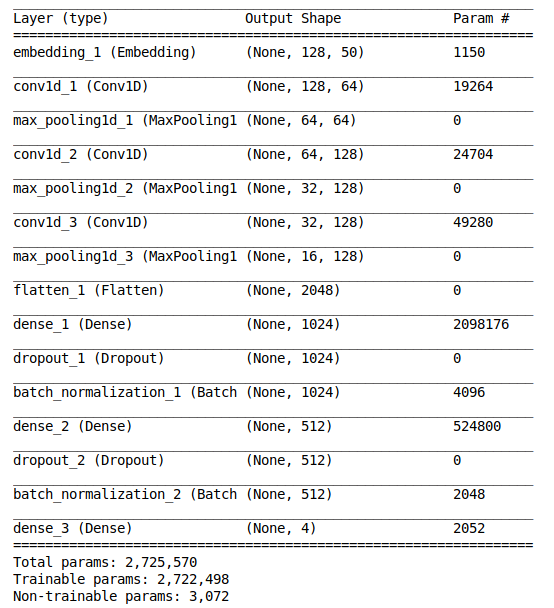}
\caption{CNN model parameters with dimensions}
\label{fig:VGGCNN}
\end{figure}

\subsection{Word Embedding}
according to Wikipedia "Word embedding is the collective name for a set of language modeling and feature learning techniques in natural language processing where words or phrases from the vocabulary are mapped to vectors of real numbers". we used two types of embedding technique for this work. the first one is word2vec ~\cite{word2vec}. the reasons we need these embedding techniques that Deep learning or machine learning model only deals with real numbers. embedding not only convert these text or sequence into number but also produce some relationship between them. we have two algorithms for word2vec, skip-gram and Common Bag Of Words (CBOW). We will not explain this algorithm here. after using both of this algorithm in all thee models we find that skip-gram work better in this case. we used the tri-gram data to generate our own word2vec model. The \ref{fig:word2vec} shows the relationship between sequences. We used different output dimension i.e 100, 150 and 300. we find that 300 dimension output has better performance.

\subsection{Convolution neural network}
Convolution neural network(CNN) is the most famous network in deep learning. we used the network architecture of VGG [~\cite{vgg}- ~\cite{snapshot}]. we are using Convolution1D, we are dealing with one dimension data here.
this network has 4 convolution layers each layer followed by max-pooling layer. the network also includes batch normalization and dropout in order to prevent the model from over-fitting. The final max-pooling layer is followed by two dense layers. 
We set the following hyper-parameters, which gives us the best results. learning rate 0.001, batch size 512, loss function cross entropy, optimizer Adam, number of epochs 20, dropout rate 0.5, activation ReLU and final layer is softmax. Figure \ref{fig:VGGCNN} shows the model architecture along with some model parameters.

\section{Experiments}
we evaluate our models on protein data bank (PDB) which is a database for the three-dimensional structural data of large biological molecules, such as proteins and nucleic acids. Performance comparison between different model will only make sense if we keep pre-processing and embedding the same. our different experiments show that word2vec with 300 dimensions has better performance and we will keep this setting unless mentioned.
\subsection{CNN Model results}
\begin{figure}[ht]
\includegraphics[width=\linewidth]{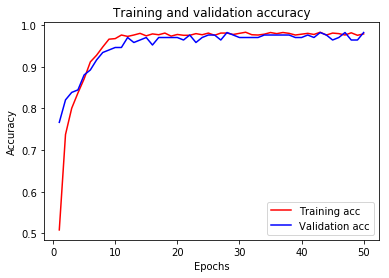}
\caption{Training and validation loss}
\label{fig:loss}
\end{figure}

\begin{figure}[ht]
\includegraphics[width=\linewidth]{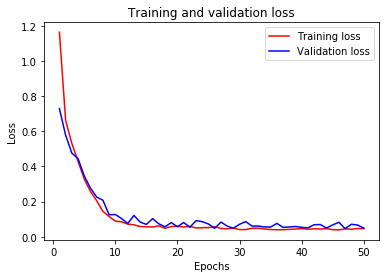}
\caption{CNN model training and validation accuracy}
\label{fig:accuracy}
\end{figure}

\begin{figure}[ht]
\includegraphics[width=\linewidth]{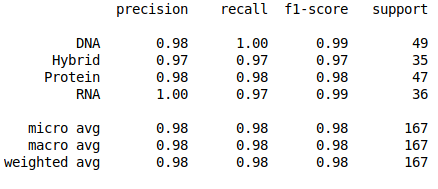}
\caption{precision, recall and F1-score}
\label{fig:performance}
\end{figure}

\begin{figure}[ht]
\includegraphics[width=\linewidth]{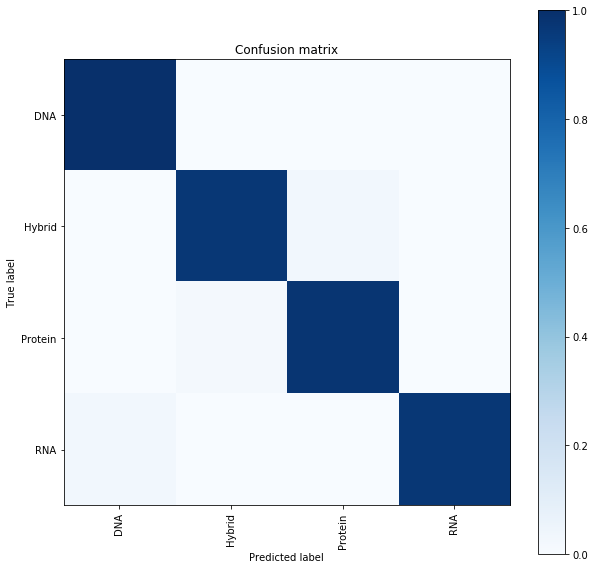}
\caption{Confusion Matrix for all classes}
\label{fig:confusion}
\end{figure}

CNN model has been explained in section IV as shown in figure \ref{fig:VGGCNN}. Embedding dimension is 50 for this model. Figure \ref{fig:loss} show the validation and training loss over the 50 epochs. we used early stopping algorithm in order to save and use our best model. Training loss is 0.028 while validation loss is 0.034. Figure \ref{fig:accuracy} show the accuracy curve for the same setting. The final training accuracy is 99.2\% and test accuracy is 98.8\%. Figure \ref{fig:performance} shows the precision, recall and F1 score of CNN model. As we can see that micro average and macro average are almost the same. Figure \ref{fig:confusion} shows the confusion matrix of all four classes. the accuracy of each class is almost the same and we can see that from diagonal color. 

\section{Additional Simulation}
Goodfellow in  ~\cite{DeepBook} explain "No free lunch theorem". In a very broad sense, it states that when averaged over all possible problems, no algorithm will perform better than all others. keeping this definition in mind we start with the simple machine learning algorithm called Random Forest. Random forest already been used in Natural language processing (NLP) and has been quite successful. the pre-processing step is the same and we use tri-gram for word2vec \ref{fig:word2vec} embedding. 


we used "gensim" to build our own word2vec model using skip-gram and we visualize the results in figure \ref{fig:word2vec}. we used tri-gram, bi-gram and uni-gram as features and average them to get single feature vector for each input sequence. the same process is repeated for testing data using same word2vec model. Random forest was initialized by 100 estimators and the results shows 96.83\% test accuracy for more than 90 thousand test sequences. we test the random forest for both balance data and unbalance data and the results was almost the same. Table \ref{table:data2} shows precision, recall and F1 score of each class. \\
compare to CNN the results are not good enough but Random forest is much faster and less expansive than CNN. This comparison may be fair but we are reporting these results to show that traditional machine learning algorithms also work better some tvisual question answeringimes. the next step is to compare different state of the art algorithm like LSTM, Random Forest and GRU with CNN. 

\begin{figure}[ht]
\includegraphics[width=\linewidth]{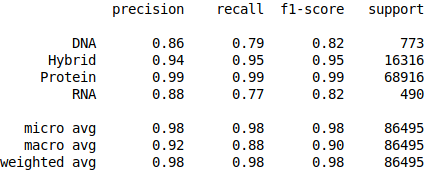}
\caption{Precison Recall for Full dataset}
\label{fig:perform1}
\end{figure}

~\cite{snapshot}
\begin{table}[ht]
\caption{Random forest results in term of precision and Recall}
\label{table:data2}
\centering
\begin{tabular}{ |c|c|c|c| } 

 \hline
 \thead{Label} & \thead{Precision} & \thead{Recall} & \thead{F-1 Score} \\ 
 
 \hline
\thead{Protein}  & \thead{0.96} & \thead{0.99} & \thead{0.97} \\ 
 
 \hline
 \thead{DNA} & \thead{0.90} & \thead{0.80} & \thead{0.85} \\ 
 
 \hline
 \thead{RNA} & \thead{0.93} & \thead{0.69} & \thead{0.79} \\
 
 \hline
 \thead{Hybrid} & \makecell{0.94} & \thead{0.88} & \thead{0.91} \\
 \hline
 \hline
\makecell{Macro Avg} & \makecell{0.93} & \thead{.84} & \thead{0.88} \\
\hline 
\makecell{Weighted avg} & \makecell{0.96} & \thead{0.96} &\thead{0.96}\\

 \hline
\end{tabular}

\end{table}
\\
Figure \ref{fig:LSTM} shows the network diagram of LSTM. we are using Tensorflow embedding in this case with 50 dimension vectors. The LSTM and GRU have the same network structure and use the same number of cells, 512 in this case. Table \ref{table:data3} shows the comparison of accuracy and loss across different network. The table clearly shows that VGG-16 ~\cite{vgg} has better performance with Word2vec ~\cite{word2vec} as embedding. seven layer CNN also perform better compare to LSTM and GRU model. All the convolution neural network are one dimension. \\
As we discussed we have full dataset and going under sampling and than using word2vec and feed it to CNN, we get the results shown in \ref{fig:perform1}.

\begin{table}[htbp]
\caption{Performance of different network}
\label{table:data3}
\centering
\begin{tabular}{ |c|c|c|c|c| } 

 \hline
 \thead{Model} & \thead{Train-Acc} & \thead{Train-Loss} & \thead{Val-Acc} & \thead{Val-Loss} \\ 
 
 \hline
\thead{CNN}  & \thead{98.19\%} & \thead{0.0486} & \thead{97.74\%} & \thead{0.0819}\\ 
 
 \hline
 \thead{GRU} & \thead{90.79\%} & \thead{0.2691} & \thead{89.70\%} & \thead{0.2715} \\ 
 
 \hline
 \thead{LSTM} & \thead{95.12\%} & \thead{0.3982} & \thead{95.14\%} & \thead{0.1962} \\
 
 \hline
 \thead{CNN-GRU} & \thead{94.85\%} & \thead{0.1509} & \thead{92.74\%} & \thead{0.1962} \\
 \hline
 \thead{RF~\cite{RandomForest}} & \thead{95.17\%} & \thead{0.4019} & \thead{94.87\%} & \thead{0.4906} \\
 
 \hline
\thead{\textbf{VGG16~\cite{vgg}}} & \textbf{99.11\%} & \textbf{0.0288} & \textbf{98.1\%} & \textbf{0.0297}\\

 \hline
\end{tabular}

\end{table}

\bibliographystyle{plain}
\bibliography{references}

\end{document}